\documentclass[11pt]{article}

\usepackage[margin=1in]{geometry}
\usepackage{amsmath,amssymb}
\usepackage{graphicx}
\usepackage{booktabs}
\usepackage{hyperref}
\usepackage{natbib}
\usepackage{xcolor}
\usepackage{microtype}
\usepackage{float}
\usepackage{caption}
\usepackage{subcaption}
\usepackage{array}
\usepackage{multirow}
\usepackage{url}
\usepackage{listings}
\usepackage{parskip}

\hypersetup{
  colorlinks=true,
  linkcolor=blue!60!black,
  citecolor=blue!60!black,
  urlcolor=blue!60!black
}

\title{\textbf{Electoral Systems Simulator}: An Open Framework for\\
Comparing Electoral Mechanisms Across Voter Distribution Scenarios}

\author{
  Sumit Mukherjee \\
  Oracle \\
  sumitmukherjee2@gmail.com
}

\date{\today}

\begin{document}

\maketitle

\begin{abstract}
Here we present \texttt{electoral\_sim}, an open-source Python framework for
simulating and comparing electoral systems across diverse voter preference
distributions. The framework represents voters and candidates as points in a
two-dimensional ideological space, derives sincere ballot profiles from
Euclidean preference distances, and evaluates several standard electoral
mechanisms---including plurality, ranked-choice, approval, score, Condorcet,
and two proportional representation systems---against a common primary
metric: the Euclidean distance between the electoral outcome and the
geometric median of the voter distribution. We evaluate these systems across
many empirically-grounded scenarios ranging from unimodal consensus
electorates to sharply polarised bimodal configurations, reporting both
single-run and Monte Carlo stability results across 200 trials per scenario.
As a case study in framework extensibility, we implement and evaluate a novel hypothetical mechanism that is not currently
implemented in any jurisdiction---in which each voter's influence is
distributed across candidates via a Boltzmann softmax kernel. This system
is included as a theoretical benchmark characterising an approximate upper bound on centroid-seeking performance, rather than as a policy proposal.
All code is released
publicly at \url{https://github.com/mukhes3/electoral_sim}.
\end{abstract}

\section{Introduction}
\label{sec:intro}

The design of electoral systems is a well studied problem in political
science and social choice theory, yet quantitative comparisons across systems
and contexts remain methodologically fragmented. Theoretical results in social
choice---Arrow's impossibility theorem \citep{arrow1951social}, the
Gibbard--Satterthwaite theorem
\citep{gibbard1973manipulation,satterthwaite1975strategy}, and the median
voter theorem \citep{black1948rationale,downs1957economic}---characterise
individual systems in isolation but do not readily yield comparative rankings
across the range of electorate configurations observed in practice. Empirical
studies, by contrast, are confounded by strategic voting, endogenous candidate
positioning, and the difficulty of observing counterfactual outcomes under alternative rules \citep{duverger1954political,lijphart2012patterns}.

Spatial models of elections \citep{hotelling1929stability,enelow1984spatial}
offer a middle ground: by representing voters and candidates as points in an ideological space, they make precise the notion of an outcome being ``close to'' voter preferences while remaining tractable enough to simulate at scale. Several simulation studies have used this framework to compare specific pairs of systems \citep{merrill1984comparison,tideman2006collective,
green2020direct}, but accessible, extensible tooling for systematic comparison across a broader set of systems and scenario types has remained limited.

This paper describes \texttt{electoral\_sim}, a Python package designed to address this gap. The primary contribution is the software framework itself: a modular architecture in which new electoral systems, electorate distributions, and evaluation metrics can be added easily. We evaluate ten systems---nine standard mechanisms and one hypothetical benchmark---across eight voter scenarios, report a primary metric grounded in
robust statistics (Section~\ref{sec:metric}), and demonstrate the framework's extensibility through the Fractional Ballot case study
(Section~\ref{sec:fractional}).

The intended audience includes political scientists and social scientists who wish to simulate how a proposed system would perform under specific electorate conditions, as well as quantitatively-minded readers from other disciplines who might be interested in learning about different electoral systems.

\section{Related Work}
\label{sec:related}

\paragraph{Simulating elections in ideological space.}
A long tradition in political science represents voters and candidates as
points in an ideological space and studies which electoral rules produce
outcomes closest to the centre of voter preferences
\citep{hotelling1929stability,downs1957economic}. Simulation studies in
this tradition---most notably \citet{merrill1984comparison} and
\citet{tideman2006collective}---have compared plurality, Borda, approval,
and Condorcet methods across a range of voter distributions, generally
finding that methods which account for the full preference ordering
outperform simple plurality in terms of proximity to the voter centre.
More recently, \citet{green2020direct} extended such comparisons to a
larger set of systems with a focus on strategy resistance. Our framework
sits in this tradition but prioritises extensibility and reproducibility:
the goal is a tool that researchers can adapt to new systems and scenarios
with minimal effort, rather than a definitive comparative study.

\paragraph{Proportional representation and legislative outcomes.}
For PR systems, the relevant outcome is not a single winner but the
composition of the resulting legislature. \citet{powell2000elections}
measures policy representation by the distance between the median voter
and the median legislator---a natural spatial analogue of proportionality
that we adopt directly as our outcome metric for PR systems.

\paragraph{Choosing a welfare benchmark.}
Any simulation framework needs a reference point against which to measure
outcomes. We use the geometric median of the voter distribution: the point
that minimises the total distance to all voters, and the natural
two-dimensional generalisation of the median voter
\citep{weiszfeld1937point}. It is more robust than the mean in polarised
electorates, where a small number of extreme voters can pull the arithmetic
average away from the true centre of mass.

\paragraph{Placing voters and politicians in ideological space.}
In our simulations, voter and candidate positions are set by design.
In practice, recovering these positions from observable data is a
non-trivial problem. Methods such as NOMINATE \citep{poole1985spatial}
infer legislator positions from roll-call voting records; Voting Advice
Applications such as Stemwijzer \citep{holleman2020voting} place voters
and parties on a common scale using structured policy questionnaires.

\section{Framework Design}
\label{sec:framework}

\subsection{Ideological Space and Preference Representation}

Voter preferences and candidate positions are represented as points in
$[0,1]^2$. Dimension~1 corresponds to an economic left--right axis;
dimension~2 to a social libertarian--authoritarian axis. The preference
distance from voter $i$ to candidate $k$ is
\begin{equation}
  d_{ik} = \|\mathbf{v}_i - \mathbf{x}_k\|_2,
\end{equation}
where $\mathbf{v}_i \in [0,1]^2$ is the voter's preference vector and
$\mathbf{x}_k \in [0,1]^2$ is the candidate's position.

\subsection{Ballot Derivation}

Ballot profiles are derived under \emph{sincere voting}: each voter ranks,
scores, and approves candidates in strict accordance with their preference
distances. Concretely:

\begin{itemize}
  \item \textbf{Plurality ballot}: vote for the nearest candidate.
  \item \textbf{Ranking}: candidates ordered by ascending $d_{ik}$.
  \item \textbf{Score ballot}: scores mapped linearly from distances,
    normalised to $[0,1]$ with the nearest candidate receiving score~1.
  \item \textbf{Approval ballot}: approve all candidates within a fixed
    distance threshold of the voter's most-preferred candidate.
\end{itemize}

All ballot types are derived once from the preference matrix and stored in a
\texttt{BallotProfile} object, ensuring all systems operate on the same
underlying data.

\subsection{Primary Evaluation Metric}
\label{sec:metric}

For a given electoral outcome $\hat{\mathbf{x}}$ and voter distribution
$\{\mathbf{v}_i\}_{i=1}^n$, the primary metric is the \emph{distance to the
geometric median}:
\begin{equation}
  \delta = \|\hat{\mathbf{x}} - \boldsymbol{\mu}^*\|_2,
  \quad \text{where} \quad
  \boldsymbol{\mu}^* = \arg\min_{\mathbf{p}} \sum_{i=1}^n
  \|\mathbf{v}_i - \mathbf{p}\|_2.
\end{equation}
The geometric median $\boldsymbol{\mu}^*$ is computed via the Weiszfeld
algorithm \citep{weiszfeld1937point}. We prefer the geometric median over the
arithmetic mean as the reference point because it is robust to outliers.

For PR systems, $\hat{\mathbf{x}}$ is taken as the \emph{median legislator
position}: the position of the candidate at the 50th percentile of the
cumulative seat-share distribution sorted by the economic axis. The seat-share centroid is also
recorded as a supplementary metric.

Additional metrics computed per system include majority satisfaction (fraction
of voters closer to the outcome than to any other candidate), mean and
worst-case voter--outcome distance, and a Gini coefficient over voter--outcome
distances as a measure of representational inequality.

\subsection{Software Architecture}

Each electoral system implements a common \texttt{ElectoralSystem} abstract
base class with a single \texttt{run(ballots, candidates) -> ElectionResult}
method. Every system returns a unified \texttt{ElectionResult} containing:
\begin{itemize}
  \item \texttt{outcome\_position}: the primary spatial outcome (winner's
    position for WTA systems; median legislator position for PR systems).
  \item \texttt{centroid\_position}: the seat-share-weighted centroid of
    elected candidates (supplementary metric for PR systems).
  \item \texttt{seat\_shares}: a dictionary mapping candidate indices to
    their share of legislative power.
  \item \texttt{winner\_indices}: list of elected candidate indices.
  \item \texttt{is\_pr}: flag indicating proportional representation.
\end{itemize}

Scenarios are defined as YAML configuration files specifying the number of
voters, candidate positions, and electorate distribution parameters. This
design allows users to specify new scenarios without modifying Python code.
An illustrative configuration is shown in Listing~\ref{lst:yaml}.

\begin{lstlisting}[caption={Excerpt from the Polarized Bimodal scenario YAML.},
  label={lst:yaml}]
name: "Polarized Bimodal"
real_world_analog: "Contemporary USA, Brexit-era UK"
n_voters: 5000
electorate:
  type: gaussian_mixture
  components:
    - weight: 0.55
      mean: [0.72, 0.58]
      std:  [0.10, 0.08]
    - weight: 0.45
      mean: [0.25, 0.38]
      std:  [0.10, 0.08]
candidates:
  - {label: "Far-Right", position: [0.80, 0.75]}
  - {label: "Right",     position: [0.72, 0.58]}
  - {label: "Center",    position: [0.50, 0.48]}
  - {label: "Left",      position: [0.28, 0.42]}
  - {label: "Far-Left",  position: [0.15, 0.25]}
\end{lstlisting}

The module structure of the package is described in
Appendix~\ref{app:architecture}.

\section{Systems Under Evaluation}
\label{sec:systems}

We evaluate ten electoral systems in total. Nine are standard mechanisms with
established theoretical and empirical literatures. The tenth---the Fractional
Ballot---is a hypothetical system introduced here as a theoretical benchmark.
It is \emph{not} currently implemented in any jurisdiction and faces
substantial practical barriers to adoption; its role in this paper is to
characterise an approximate upper bound on the centroid-seeking performance
achievable by an electoral mechanism. Its definition is given in
Section~\ref{sec:fractional}; the practical challenges it poses are discussed
in Section~\ref{sec:discussion}.

\subsection{Winner-Take-All Systems}

Let $n$ denote the number of voters, $K$ the number of candidates, and
$r_{ik} \in \{1, \ldots, K\}$ the rank assigned by voter $i$ to candidate
$k$ (rank 1 = most preferred). Let $s_{ik} \in [0,1]$ denote the score
assigned by voter $i$ to candidate $k$, and $a_{ik} \in \{0,1\}$ the
approval indicator.

\begin{enumerate}

  \item \textbf{Plurality (FPTP)}: each voter casts one vote for their
    top-ranked candidate. The winner is
    \begin{equation}
      k^* = \arg\max_k \sum_{i=1}^n \mathbf{1}[r_{ik} = 1].
    \end{equation}

  \item \textbf{Two-Round Runoff}: let $v_k = \sum_i \mathbf{1}[r_{ik}=1]$
    be the first-round vote totals. If $\max_k v_k > n/2$ the plurality
    winner is elected outright. Otherwise the two candidates with the
    highest $v_k$ advance to a runoff, and the candidate preferred by more
    voters wins:
    \begin{equation}
      k^* = \arg\max_{k \,\in\, \{k_1, k_2\}} \sum_{i=1}^n
      \mathbf{1}[r_{ik} < r_{ik'}], \quad k' \neq k.
    \end{equation}

  \item \textbf{Instant Runoff Voting (IRV)} \citep{tideman1987independence}:
    candidates are eliminated iteratively. At each round the candidate with
    the fewest first-preference votes among remaining candidates is
    eliminated and their votes redistributed to the next-ranked remaining
    candidate. Formally, let $\mathcal{C}^{(t)}$ be the candidate set at
    round $t$ and $v_k^{(t)}$ the corresponding first-preference totals.
    The eliminated candidate is
    \begin{equation}
      e^{(t)} = \arg\min_{k \in \mathcal{C}^{(t)}} v_k^{(t)},
    \end{equation}
    and the process repeats until $|\mathcal{C}^{(t)}| = 1$.

  \item \textbf{Borda Count} \citep{borda1781memoire}: each candidate
    receives $K - r_{ik}$ points from voter $i$. The winner maximises the
    total Borda score:
    \begin{equation}
      k^* = \arg\max_k \sum_{i=1}^n (K - r_{ik}).
    \end{equation}

  \item \textbf{Approval Voting} \citep{brams1978approval}: the winner
    maximises total approvals:
    \begin{equation}
      k^* = \arg\max_k \sum_{i=1}^n a_{ik}.
    \end{equation}
    In our implementation, voter $i$ approves all candidates within a
    distance threshold $\tau$ of their most-preferred candidate:
    $a_{ik} = \mathbf{1}[d_{ik} \leq \tau \cdot d_{i,\text{min}}]$, where
    $d_{i,\text{min}} = \min_j d_{ij}$.

  \item \textbf{Score Voting}: the winner maximises the mean score across
    voters:
    \begin{equation}
      k^* = \arg\max_k \frac{1}{n} \sum_{i=1}^n s_{ik}.
    \end{equation}

  \item \textbf{Condorcet--Schulze} \citep{schulze2011new}: the method
  proceeds in two stages.

  \textit{Stage 1 --- pairwise comparisons.} For each pair of candidates
  $(k, k')$, count how many voters prefer $k$ over $k'$:
  \begin{equation}
    d(k, k') = \sum_{i=1}^n \mathbf{1}[r_{ik} < r_{ik'}].
  \end{equation}
  Candidate $k$ \emph{beats} $k'$ directly if $d(k, k') > n/2$.
  If one candidate beats all others directly, they are the Condorcet
  winner and are elected immediately.

  \textit{Stage 2 --- beatpath (Schulze).} When no Condorcet winner
  exists, the method considers indirect dominance. A beatpath from $k$
  to $k'$ is any sequence of candidates $k \to c_1 \to c_2 \to \cdots
  \to k'$ where each step is a direct win. The strength of a beatpath
  is the weakest margin along it, and the strongest beatpath strength is
  \begin{equation}
    p(k, k') = \max_{\text{paths } k \to k'}
    \min_{\text{edges on path}} d(\cdot, \cdot).
  \end{equation}
  The winner is the candidate $k^*$ who has a stronger beatpath to every
  other candidate than they have back:
  \begin{equation}
    k^* : \quad p(k^*, k') \geq p(k', k^*) \quad \forall\, k' \neq k^*.
  \end{equation}

\end{enumerate}

\subsection{Proportional Representation Systems}

Let $S$ denote the total number of seats (we use $S = 100$ throughout).

\begin{enumerate}
  \setcounter{enumi}{7}

  \item \textbf{Party-List PR (D'Hondt)} \citep{dhondt1882systeme}: seats
    are awarded one at a time. At each step, every party $k$ is assigned
    the quotient $v_k / (s_k + 1)$, where $v_k$ is its vote share and
    $s_k$ is the number of seats it has received so far (initially zero).
    The party with the highest quotient receives the next seat, and the
    process repeats until all $S$ seats are filled:
    \begin{equation}
      \text{award seat to } \arg\max_k \frac{v_k}{s_k + 1}.
    \end{equation}

  \item \textbf{Mixed Member Proportional (MMP)}: seat allocation proceeds
    in two stages. First, $\lfloor S/2 \rfloor$ \emph{district seats} are
    filled by plurality winners in single-member districts, giving each
    party $s_k^{\text{district}}$ seats. Second, the remaining
    $\lceil S/2 \rceil$ \emph{list seats} are distributed to bring each
    party's total as close as possible to its proportional entitlement
    $v_k \cdot S$:
    \begin{equation}
      s_k^{\text{list}} = \max\!\left(0,\;
      \lfloor v_k \cdot S \rfloor - s_k^{\text{district}} \right).
    \end{equation}
    Parties that won more district seats than their proportional share
    retain the excess (an overhang); their list allocation is zero.

\end{enumerate}

\subsection{Hypothetical Benchmark}

\begin{enumerate}
  \setcounter{enumi}{9}
  \item \textbf{Fractional Ballot} : each voter's influence is
    distributed across candidates via a Boltzmann softmax kernel
    parameterised by temperature $\sigma$. We evaluate three values
    $\sigma \in \{0.1, 0.3, 1.0\}$ in both a single-winner (Discrete) and a
    weighted-legislature (Continuous) variant. Full definition in
    Section~\ref{sec:fractional}.
\end{enumerate}

\section{Scenarios}
\label{sec:scenarios}

We define eight scenarios intended to represent distinct real-world electorate
configurations (Table~\ref{tab:scenarios}). The five general scenarios vary
the shape of the voter distribution; the three two-party scenarios test
systems under conditions analogous to formal primary processes.

\begin{table}[h]
\centering
\caption{Scenario descriptions. Voters are sampled from parameterised
Gaussian mixture models. The geometric median--mean gap
$\|\boldsymbol{\mu}^* - \bar{\mathbf{v}}\|$ reflects distributional
asymmetry.}
\label{tab:scenarios}
\begin{tabular}{llll}
\toprule
\# & Name & Real-world analogue & Voters / Candidates \\
\midrule
1 & Unimodal Consensus          & Nordic / consensus democracies         & 5{,}000 / 7 \\
2 & Polarized Bimodal           & Contemporary USA, Brexit-era UK        & 5{,}000 / 5 \\
3 & Multimodal Fragmented       & Netherlands, Israel, Italy             & 5{,}000 / 8 \\
4 & Dominant Party              & Japan (LDP), Hungary                   & 5{,}000 / 6 \\
5 & Asymmetric Skewed           & Latin America, post-Soviet states      & 5{,}000 / 6 \\
6 & Two-Party Symmetric         & Stylised US two-party system           & 6{,}000 / 6 \\
7 & Two-Party Centrist Majority & Many European two-bloc systems         & 6{,}000 / 6 \\
8 & Two-Party Dominant Left     & Dominant-party with primary process    & 6{,}000 / 7 \\
\bottomrule
\end{tabular}
\end{table}

\begin{figure}[H]
  \centering
  \includegraphics[width=\textwidth]{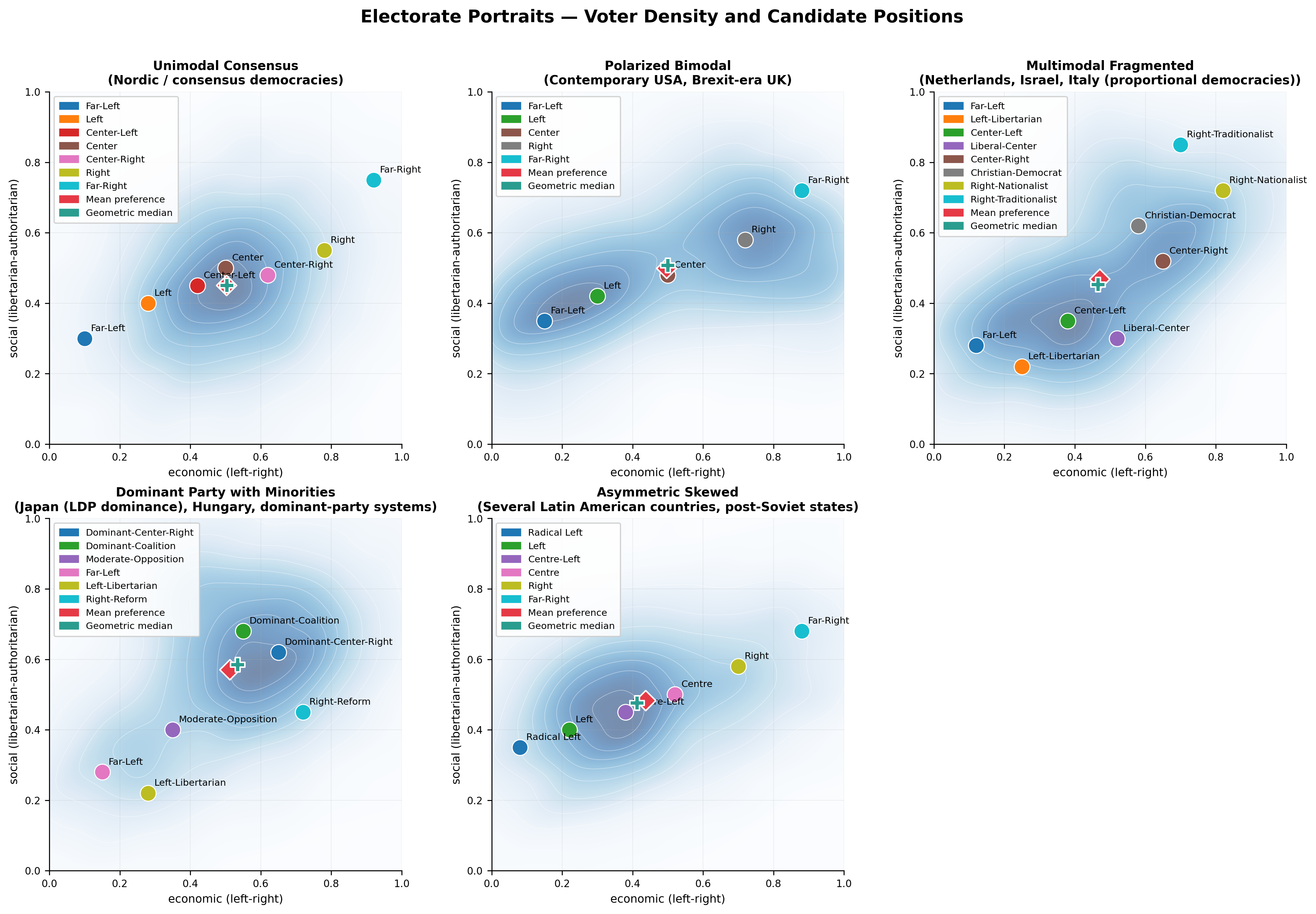}
  \caption{Electorate portraits for the five general scenarios. Blue contours
    show kernel-density estimates of voter preferences. The teal cross
    ($\boldsymbol{+}$) marks the geometric median; the red diamond
    ($\blacklozenge$) marks the arithmetic mean. Candidate positions are
    labelled. The ideal electoral outcome minimises distance to the teal
    marker.}
  \label{fig:portraits}
\end{figure}

Electorate portraits for the five general scenarios are shown in
Figure~\ref{fig:portraits}. In the Unimodal Consensus scenario the geometric
median and arithmetic mean nearly coincide ($\Delta = 0.0006$), so any system
identifying the modal candidate performs well. The Polarized Bimodal scenario
is the most diagnostically demanding: the geometric median lies in a
low-density gap between two voter clusters ($\Delta = 0.0096$ between median
and mean), and systems anchored to plurality logic are captured by the larger
cluster rather than the cross-cluster centrist position.

\section{Results}
\label{sec:results}

\subsection{Cross-Scenario Comparison}

\begin{figure}[H]
  \centering
  \includegraphics[width=\textwidth]{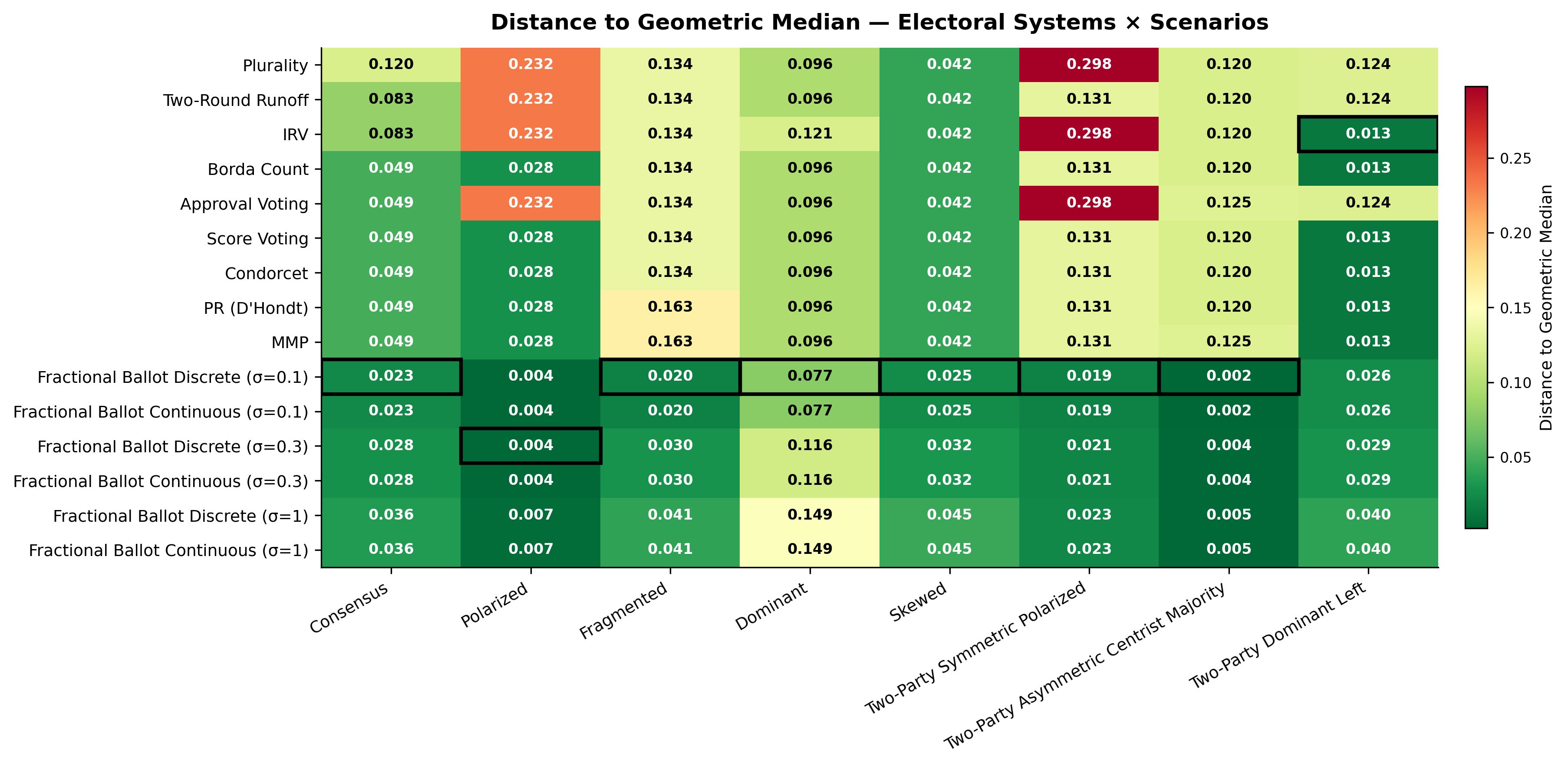}
  \caption{Distance to geometric median for all electoral systems and
    scenarios. Lower values (greener cells) indicate outcomes closer to the
    population geometric median. Black borders highlight the best-performing
    system within each scenario column. Fractional Ballot rows are labelled
    FB Discrete / Continuous; see Section~\ref{sec:fractional} for their
    definition.}
  \label{fig:heatmap}
\end{figure}

Figure~\ref{fig:heatmap} shows the primary metric---distance to the geometric
median---for all 15 system configurations (nine standard systems plus six
Fractional Ballot variants) across all eight scenarios.

\subsubsection{Standard Systems}

Several patterns emerge consistently across scenarios.

\paragraph{Plurality performs worst or near-worst in polarised settings.}
In the Polarized Bimodal scenario, Plurality yields $\delta = 0.2321$,
representing a 53.8-fold increase relative to the best-performing system.
Two-Round Runoff and IRV produce identical outcomes ($\delta = 0.2321$):
both collapse to electing the largest-cluster candidate, since no round of
vote-counting transfers sufficient weight to the cross-cluster centrist
position. Approval Voting also fails here---the majority cluster's candidates
receive approval from enough voters to win outright.

\paragraph{Condorcet, Score, and Borda are competitive in non-polarised scenarios.}
In the Unimodal Consensus, Asymmetric Skewed, and Dominant Party scenarios,
Borda Count, Score Voting, and the Condorcet--Schulze method produce
identical outcomes ($\delta = 0.049$, $0.042$, and $0.096$ respectively),
all superior to Plurality. This convergence is consistent with the
well-established result that these methods agree when a Condorcet winner
exists and preferences are approximately single-peaked
\citep{merrill1984comparison}.

\paragraph{PR systems carry a median-legislator artefact in fragmented scenarios.}
In the Multimodal Fragmented scenario, the D'Hondt and MMP seat-share
\emph{centroids} are close to the geometric median ($\delta_\text{centroid}
= 0.0148$ and $0.0099$ respectively), but the \emph{median legislator
positions} diverge substantially ($\delta = 0.1632$ for both), producing
artefact gaps of $+0.148$ and $+0.153$. This reflects a structural property
of PR in fragmented legislatures: when seats are distributed across many
ideologically distant parties, the 50th-percentile legislator need not lie
near the voter-distribution centre. We report both metrics and recommend the
centroid as the more informative outcome proxy for PR systems in fragmented
settings.

\paragraph{Performance in the Dominant Party scenario is heterogeneous.}
This is the only scenario in which standard plurality-based systems
($\delta \approx 0.096$) and the Fractional Ballot at low $\sigma$
($\delta = 0.078$) are roughly comparable. We return to this in
Section~\ref{sec:fractional}.

\subsubsection{Fractional Ballot}

The Fractional Ballot variants consistently achieve the lowest $\delta$
across most scenarios. At $\sigma = 0.1$ the system achieves the best result
in six of eight scenarios; at $\sigma = 0.3$ it wins the Polarized Bimodal
scenario. Compared to the best standard system, the improvement is most
substantial in the Polarized Bimodal (0.0043 vs.\ 0.0278, a 6.5-fold
reduction) and Multimodal Fragmented cases (0.0197 vs.\ 0.1344, a 6.8-fold
reduction). These results are discussed in detail in
Section~\ref{sec:fractional}.

\subsection{Per-Scenario Spatial Outcomes}

\begin{figure}[H]
  \centering
  \includegraphics[width=\textwidth]{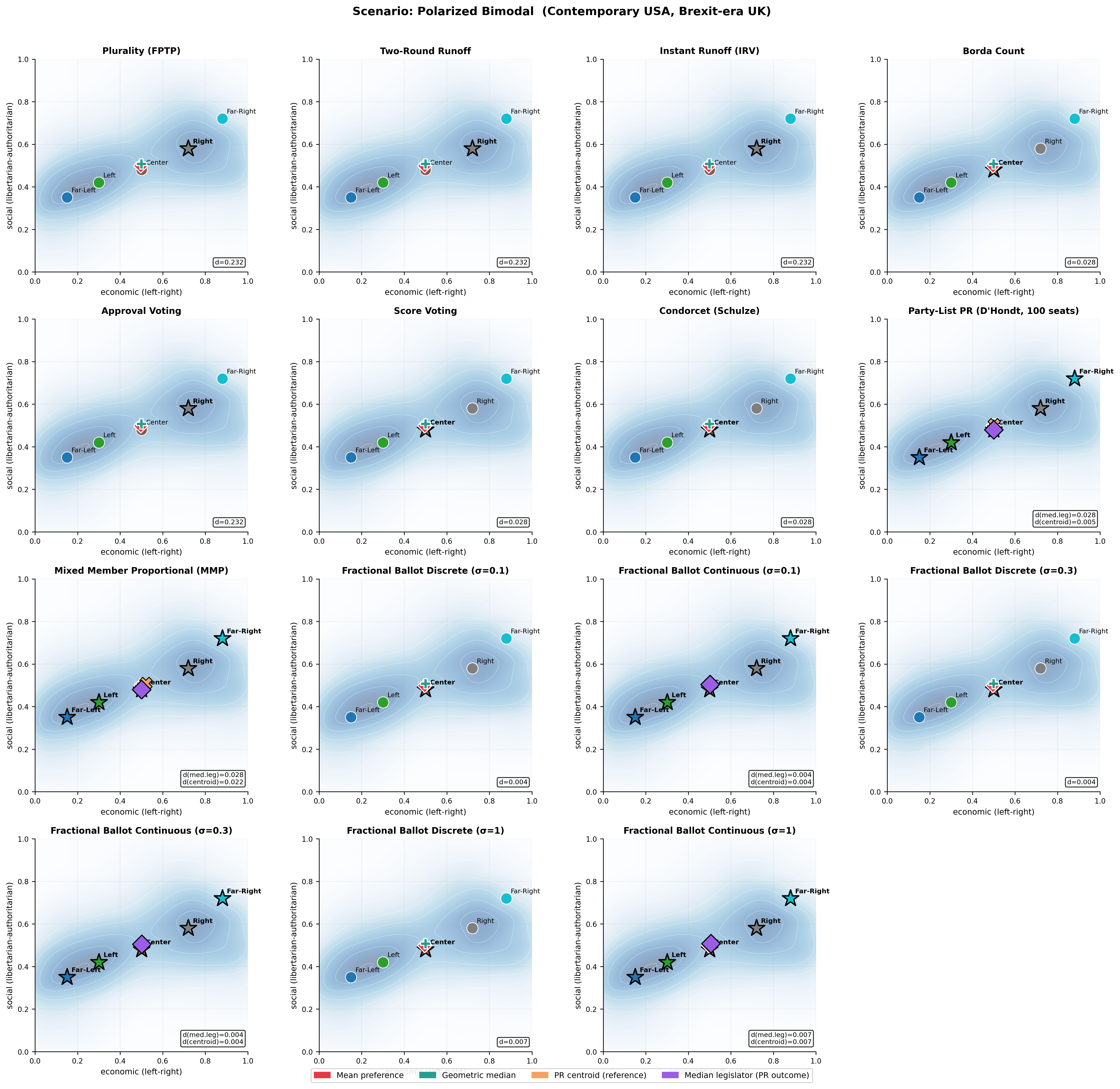}
  \caption{Per-system spatial outcome maps for the Polarized Bimodal
    scenario. The teal cross marks the geometric median; the star
    ($\bigstar$) marks the elected candidate. For PR systems, the cross
    ($\times$) marks the seat-share centroid and the purple diamond marks
    the median legislator. Distance to the geometric median ($\delta$) is
    annotated in each panel.}
  \label{fig:spatial_polarized}
\end{figure}

Figure~\ref{fig:spatial_polarized} shows per-system spatial outcome maps for
the Polarized Bimodal scenario, the most discriminating of the eight
configurations.

\subsection{Monte Carlo Stability}

\begin{figure}[H]
  \centering
  \includegraphics[width=\textwidth]{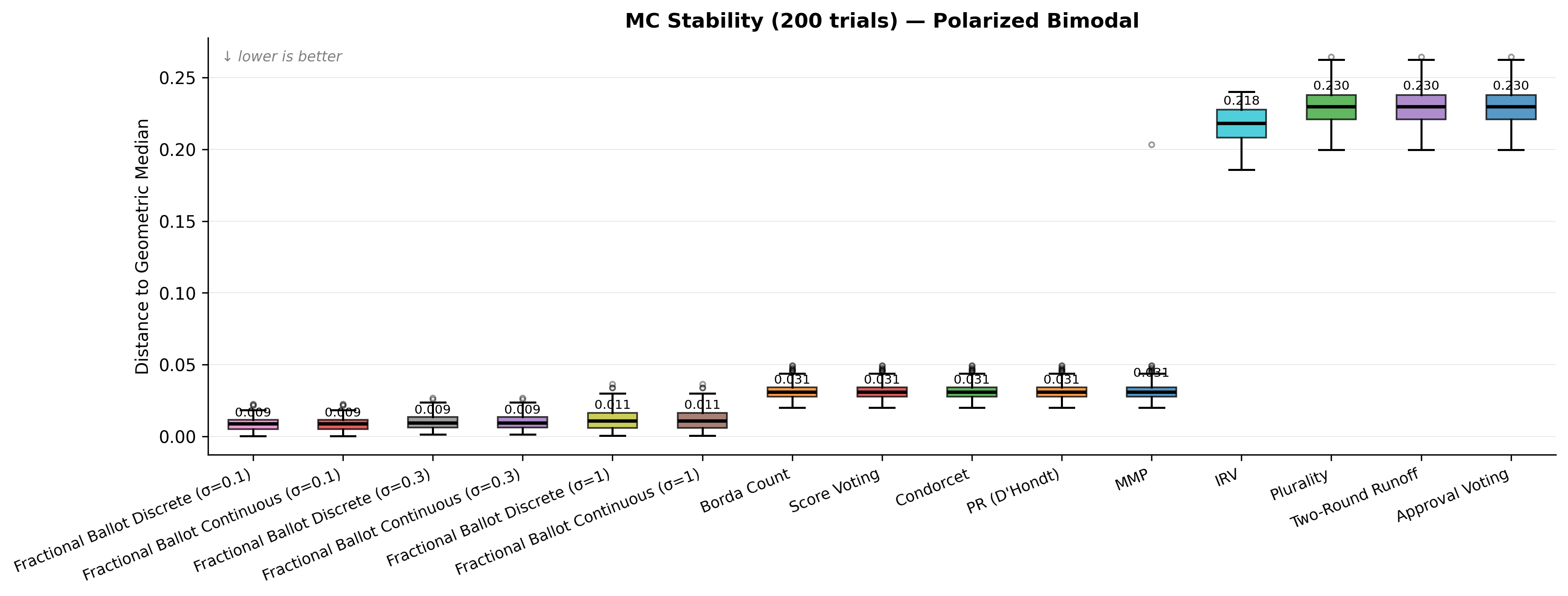}
  \caption{Monte Carlo stability: distribution of distance to geometric
    median across 200 trials for the Polarized Bimodal scenario. Boxes show
    interquartile range; whiskers extend to the 5th and 95th percentiles.
    Systems are ordered left-to-right by median $\delta$.}
  \label{fig:mc}
\end{figure}

To assess sensitivity to the specific random draw of the electorate, we ran
200 trials per scenario, each re-sampling $n = 2{,}000$ voters from the same
distribution. Figure~\ref{fig:mc} shows the distribution of $\delta$ across
trials for the Polarized Bimodal scenario; results for all scenarios are
available in the supplementary notebook.

The ranking of systems is stable across all 200 trials in every scenario.
The variance of $\delta$ for centroid-based methods (including the Fractional
Ballot) is substantially lower than for plurality-based methods, particularly
in the Polarized Bimodal scenario where the plurality outcome is sensitive to
which cluster slightly outnumbers the other in any given draw.

\section{The Fractional Ballot: Definition and Analysis}
\label{sec:fractional}

\subsection{Definition}

Each voter $i$ is assigned a weight vector $\mathbf{w}_i \in \Delta^{K-1}$
over $K$ candidates via a Boltzmann (softmax) kernel:
\begin{equation}
  w_{ik} = \frac{\exp(-d_{ik}/\sigma)}{\sum_{j=1}^{K} \exp(-d_{ij}/\sigma)},
  \label{eq:boltzmann}
\end{equation}
where $\sigma > 0$ is a temperature parameter controlling weight
concentration. The population mean weight vector is
\begin{equation}
  \bar{\mathbf{w}} = \frac{1}{n}\sum_{i=1}^n \mathbf{w}_i,
  \label{eq:meanw}
\end{equation}
and the electoral centroid is $\hat{\mathbf{x}} = \bar{\mathbf{w}}^\top
\mathbf{X}$, where $\mathbf{X} \in \mathbb{R}^{K \times 2}$ is the matrix of
candidate positions. We implement two variants:

\begin{itemize}
  \item \textbf{Fractional Ballot Discrete (FBD)}: a single candidate wins
    outright---the one nearest to $\hat{\mathbf{x}}$---but the
    \texttt{outcome\_position} used for metric computation is the centroid
    $\hat{\mathbf{x}}$ itself. Intended for single-winner contexts such as
    executive elections.

  \item \textbf{Fractional Ballot Continuous (FBC)}: each candidate $k$
    holds fractional legislative power $\bar{w}_k$; the outcome position is
    $\hat{\mathbf{x}}$ directly. Intended for weighted-legislature or
    policy-voting contexts.
\end{itemize}

Since both variants use $\hat{\mathbf{x}}$ as the primary outcome position
for metric computation, they produce identical $\delta$ values across all
scenarios. Table~\ref{tab:fractional} therefore reports results for the
Discrete variant only; Continuous results are identical in every cell.
The distinction between the variants is one of institutional interpretation
rather than spatial performance: FBD is appropriate for single-winner
contexts, FBC for weighted-legislature settings.

\paragraph{Limiting behaviour.}
As $\sigma \to 0$, weights concentrate on the nearest candidate and the
system converges to Plurality. As $\sigma \to \infty$, weights become
uniform ($\bar{w}_k \to 1/K$) and the centroid approaches the equal-weight
mean of candidate positions. At intermediate $\sigma$, the centroid
approximates the geometric median of the voter distribution.

\subsection{Fractional Ballot Results}

Table~\ref{tab:fractional} summarises $\delta$ for all Fractional Ballot
variants alongside the best-performing standard system in each scenario.

\begin{table}[H]
\centering
\caption{Distance to geometric median: Fractional Ballot Discrete (FBD)
versus best standard system per scenario. Lower is better. Bold entries
indicate the best result in each column. Fractional Ballot Continuous
results are identical to FBD in all cells; see Section~\ref{sec:fractional}.}
\label{tab:fractional}
\small
\begin{tabular}{lcccccc}
\toprule
System & Consensus & Polarized & Fragmented & Dominant & Skewed & Sym.\ Pol. \\
\midrule
FBD ($\sigma{=}0.1$) & \textbf{0.0235} & 0.0044          & \textbf{0.0197} & \textbf{0.0775} & \textbf{0.0247} & \textbf{0.0187} \\
FBD ($\sigma{=}0.3$) & 0.0281          & \textbf{0.0043} & 0.0301          & 0.1158          & 0.0321          & 0.0213          \\
FBD ($\sigma{=}1.0$) & 0.0357          & 0.0069          & 0.0409          & 0.1486          & 0.0446          & 0.0228          \\
\midrule
Best standard system & 0.0492          & 0.0278          & 0.1344          & 0.0962          & 0.0417          & 0.1309          \\
\bottomrule
\end{tabular}
\end{table}

\paragraph{The Dominant Party exception.}
The Dominant Party scenario is the principal failure mode of the Fractional
Ballot. The dominant coalition occupies a position well removed from the
geometric median in the authoritarian quadrant; Boltzmann weighting at
$\sigma \geq 0.3$ is pulled sufficiently toward this large cluster that
$\delta$ substantially exceeds that of standard systems. The $\sigma = 0.1$ variant partially
mitigates this, yielding $\delta = 0.0775$, but does not match the best
standard systems. This is an inherent limitation of centroid-based methods
when the largest voter cluster is spatially distant from the geometric
median: the population centroid is dragged toward the greater mass.

\paragraph{The Two-Party Dominant Left exception.}
In this scenario IRV achieves $\delta = 0.0134$ while FBD at $\sigma = 0.1$
yields $\delta = 0.0257$. Inspection of the spatial maps reveals that the
dominant left party's candidate positions bracket the geometric median
closely, so IRV's sequential elimination converges to one of them. This is a
candidate-placement coincidence rather than a structural advantage for IRV.

\section{Discussion}
\label{sec:discussion}

\subsection{Limitations of the Simulation}

\paragraph{Voters always vote honestly.}
In our simulation, every voter votes for the candidate closest to their
own preferences. In practice, voters sometimes vote strategically---for
example, supporting a second-choice candidate they believe can win rather
than a first-choice candidate they believe cannot. This is common under
Plurality voting, where voters often feel pressure to avoid ``wasting''
their vote \citep{duverger1954political}. Our results therefore reflect
a best-case version of each system; real-world outcomes under strategic
voting may differ.

\paragraph{Politics is two-dimensional here.}
We represent voter and candidate preferences on two axes: economic
left--right and social libertarian--authoritarian. Real electorates have
more dimensions---healthcare, foreign policy, immigration, and so on. Our
results may not fully generalise to settings where additional issue
dimensions matter.

\paragraph{Measuring PR outcomes is imperfect.}
For proportional representation systems, we measure the outcome as the
position of the median legislator in the resulting parliament. In
fragmented parliaments with many parties, this can be a poor summary of
where policy will actually land, since government formation and coalition
bargaining determine which parties hold power. A more realistic model
would incorporate coalition dynamics, which we leave for future work.

\subsection{Practical Implementation of the Fractional Ballot}
\label{sec:practical}

The Fractional Ballot requires that voters and candidates be located on a
common ideological scale before weights can be computed---a prerequisite
that does not arise for any standard system, where preferences are revealed
implicitly through the ballot itself. Voting Advice Applications (VAAs)
such as Stemwijzer \citep{holleman2020voting}, which already compute
voter--candidate distances from structured policy questionnaires, offer the
most natural approximation: candidates and voters complete the same
questionnaire, distances $d_{ik}$ are derived from the responses, and
weights follow from equation~\eqref{eq:boltzmann}. Established scaling
methods such as NOMINATE \citep{poole1985spatial} and manifesto-based
approaches \citep{budge2001mapping} provide complementary tools for
positioning incumbents retrospectively. Residual challenges include the
choice of dimensionality reduction, potential strategic responses to a
preference-eliciting questionnaire, and the calibration of $\sigma$. The
Fractional Ballot is therefore best understood as a theoretical benchmark
motivating further investigation into preference-eliciting mechanisms
rather than an immediately deployable system.

\subsection{Framework Extensibility}

New electoral systems can be added by subclassing \texttt{ElectoralSystem}
and implementing the \texttt{run} method (typically 20--40 lines of Python).
New scenarios require only a YAML configuration file. New metrics are added
to the \texttt{ElectionMetrics} dataclass. This modularity is the primary
design goal of the framework and is intended to minimise the overhead of
``what-if'' explorations for researchers without deep software engineering
backgrounds.

\section{Conclusion}
\label{sec:conclusion}

We have described \texttt{electoral\_sim}, an open-source Python framework
for the simulation and comparison of electoral systems across diverse voter
preference distributions. The framework implements nine standard electoral
mechanisms alongside one hypothetical benchmark, evaluates outcomes against
the geometric median of the voter distribution across eight
empirically-grounded scenarios, and supports straightforward extension with
new systems, scenarios, and metrics.

The simulation results confirm several theoretical predictions---Plurality
fails badly under polarisation, Condorcet methods are competitive where
approximately single-peaked preferences hold, and PR systems' median
legislator positions diverge from the voter-distribution centre in fragmented
settings---while providing a quantitative cross-scenario ranking that is
difficult to obtain from theory alone.

The Fractional Ballot case study illustrates the framework's extensibility
and provides an approximate upper bound on centroid-seeking performance. Its
near-optimal behaviour in most scenarios, and its interpretable failure mode
in the Dominant Party scenario, establish the geometric median of the voter
distribution as a productive benchmark for evaluating how well practical
systems approximate the spatial optimum. The connections between the
Fractional Ballot and existing preference-elicitation infrastructure suggest
a concrete research agenda for its practical approximation.

We hope the framework is useful to researchers and practitioners interested
in exploring how electoral design choices interact with electorate structure,
and invite contributions of new systems and scenarios via the public
repository.

\bibliographystyle{plainnat}
\bibliography{references}

@book{arrow1951social,
  author    = {Arrow, Kenneth J.},
  title     = {Social Choice and Individual Values},
  publisher = {Wiley},
  address   = {New York},
  year      = {1951}
}

@article{gibbard1973manipulation,
  author    = {Gibbard, Allan},
  title     = {Manipulation of Voting Schemes: A General Result},
  journal   = {Econometrica},
  volume    = {41},
  number    = {4},
  pages     = {587--601},
  year      = {1973}
}

@article{satterthwaite1975strategy,
  author    = {Satterthwaite, Mark A.},
  title     = {Strategy-Proofness and {A}rrow's Conditions: Existence and
               Correspondence Theorems for Voting Procedures and Social
               Welfare Functions},
  journal   = {Journal of Economic Theory},
  volume    = {10},
  number    = {2},
  pages     = {187--217},
  year      = {1975}
}

@article{black1948rationale,
  author    = {Black, Duncan},
  title     = {On the Rationale of Group Decision-Making},
  journal   = {Journal of Political Economy},
  volume    = {56},
  number    = {1},
  pages     = {23--34},
  year      = {1948}
}

@book{downs1957economic,
  author    = {Downs, Anthony},
  title     = {An Economic Theory of Democracy},
  publisher = {Harper \& Row},
  address   = {New York},
  year      = {1957}
}

@article{hotelling1929stability,
  author    = {Hotelling, Harold},
  title     = {Stability in Competition},
  journal   = {The Economic Journal},
  volume    = {39},
  number    = {153},
  pages     = {41--57},
  year      = {1929}
}

@book{enelow1984spatial,
  author    = {Enelow, James M. and Hinich, Melvin J.},
  title     = {The Spatial Theory of Voting: An Introduction},
  publisher = {Cambridge University Press},
  address   = {Cambridge},
  year      = {1984}
}

@article{merrill1984comparison,
  author    = {Merrill, Samuel},
  title     = {A Comparison of Efficiency of Multicandidate Electoral Systems},
  journal   = {American Journal of Political Science},
  volume    = {28},
  number    = {1},
  pages     = {23--48},
  year      = {1984}
}

@book{tideman2006collective,
  author    = {Tideman, Nicolaus},
  title     = {Collective Decisions and Voting: The Potential for Public Choice},
  publisher = {Ashgate},
  address   = {Aldershot},
  year      = {2006}
}

@article{green2020direct,
  title={Statistical evaluation of voting rules},
  author={Green-Armytage, James and Tideman, T Nicolaus and Cosman, Rafael},
  journal={Social Choice and Welfare},
  volume={46},
  number={1},
  pages={183--212},
  year={2016},
  publisher={Springer}
}

@book{duverger1954political,
  author    = {Duverger, Maurice},
  title     = {Political Parties: Their Organization and Activity in the
               Modern State},
  publisher = {Wiley},
  address   = {New York},
  year      = {1954},
  note      = {Translated by Barbara and Robert North}
}

@book{lijphart2012patterns,
  author    = {Lijphart, Arend},
  title     = {Patterns of Democracy: Government Forms and Performance in
               Thirty-Six Countries},
  edition   = {2nd},
  publisher = {Yale University Press},
  address   = {New Haven},
  year      = {2012}
}

@book{powell2000elections,
  author    = {Powell, G. Bingham},
  title     = {Elections as Instruments of Democracy: Majoritarian and
               Proportional Visions},
  publisher = {Yale University Press},
  address   = {New Haven},
  year      = {2000}
}

@article{borda1781memoire,
  author    = {de Borda, Jean-Charles},
  title     = {M{\'{e}}moire sur les {\'{e}}lections au scrutin},
  journal   = {M{\'{e}}moires de l'Acad{\'{e}}mie Royale des Sciences},
  year      = {1781},
  note      = {Reprinted and translated in \textit{Iain McLean and Arnold B.
               Urken (eds.), Classics of Social Choice}, University of
               Michigan Press, 1995}
}

@book{brams1978approval,
  author    = {Brams, Steven J. and Fishburn, Peter C.},
  title     = {Approval Voting},
  publisher = {Birkh{\"{a}}user},
  address   = {Boston},
  year      = {1983}
}

@article{schulze2011new,
  author    = {Schulze, Markus},
  title     = {A New Monotonic, Clone-Independent, Reversal Symmetric, and
               Condorcet-Consistent Single-Winner Election Method},
  journal   = {Social Choice and Welfare},
  volume    = {36},
  number    = {2},
  pages     = {267--303},
  year      = {2011}
}

@article{tideman1987independence,
  author    = {Tideman, T. Nicolaus},
  title     = {Independence of Clones as a Criterion for Voting Rules},
  journal   = {Social Choice and Welfare},
  volume    = {4},
  number    = {3},
  pages     = {185--206},
  year      = {1987}
}

@book{dhondt1882systeme,
  author    = {d'Hondt, Victor},
  title     = {Syst\`{e}me pratique et raisonn\'{e} de repr\'{e}sentation
               proportionnelle},
  publisher = {Muquardt},
  address   = {Brussels},
  year      = {1882}
}

@article{weiszfeld1937point,
  author    = {Weiszfeld, Endre},
  title     = {Sur le point pour lequel la somme des distances de $n$ points
               donn\'{e}s est minimum},
  journal   = {T\^{o}hoku Mathematical Journal},
  volume    = {43},
  pages     = {355--386},
  year      = {1937}
}

@book{poole1985spatial,
  author    = {Poole, Keith T. and Rosenthal, Howard},
  title     = {Congress: A Political-Economic History of Roll Call Voting},
  publisher = {Oxford University Press},
  address   = {New York},
  year      = {1997}
}

@book{budge2001mapping,
  author    = {Budge, Ian and Klingemann, Hans-Dieter and Volkens, Andrea
               and Bara, Judith and Tanenbaum, Eric},
  title     = {Mapping Policy Preferences: Estimates for Parties, Electors,
               and Governments 1945--1998},
  publisher = {Oxford University Press},
  address   = {Oxford},
  year      = {2001}
}

@article{holleman2020voting,
  author    = {Holleman, Bouwe and Kamoen, Naomi and Krouwel, Andr\'{e}
               and de Vreese, Claes and van de Pol, Jante},
  title     = {Voting Advice Applications and Electoral Outcomes},
  journal   = {Political Communication},
  volume    = {37},
  number    = {3},
  pages     = {394--413},
  year      = {2020}
}

\appendix

\section{Software Architecture}
\label{app:architecture}

\begin{lstlisting}[caption={Module structure of \texttt{electoral\_sim}.}]
electoral_sim/
  types.py           # ElectionResult dataclass
  scenario.py        # YAML loader: load_scenario, load_all_scenarios
  fractional.py      # Fractional Ballot variants
  electorate/        # Electorate class, distribution factories
  candidates/        # CandidateSet class
  ballots/           # BallotProfile, sincere ballot derivation
  systems/           # ElectoralSystem ABC + 9 implementations
  metrics/           # ElectionMetrics, run_simulation, run_monte_carlo
  primaries/         # Two-party primary pipeline
  utils/
    viz_electorate.py
    viz_metrics.py
\end{lstlisting}

\section{Reproducibility}
\label{app:repro}

All results in this paper are produced by the Jupyter notebook\\
\texttt{notebooks/01\_electoral\_systems\_comparison.ipynb} included in the
repository. The notebook uses a fixed random seed (\texttt{SEED=42}) for all
stochastic operations. 

\end{document}